# Pressure-induced superconductivity in TiGeTe$_6$

Sayaka Yamamoto[a,b], *Ryo Matsumoto[c], Shintaro Adachi[d], Kensei Terashima[a], Hiromi Tanaka[e,f] Tetsuo Irifune[g], Hiroyuki Takeya[a], and Yoshihiko Takano[a,b]

[a]*International Center for Materials Nanoarchitectonics (MANA), National Institute for Materials Science, Tsukuba, Ibaraki 305-0047, Japan*
[b]*Graduate School of Pure and Applied Sciences, University of Tsukuba, 1-1-1 Tennodai, Tsukuba, Ibaraki 305-8577, Japan*
[c]*International Center for Young Scientists (ICYS), National Institute for Materials Science, Tsukuba, Ibaraki 305-0047, Japan*
[d]*Nagamori Institute of Actuators, Kyoto University of Advanced Science, Ukyo-ku, Kyoto 615-8577, Japan*
[e]*National Institute of Technology, Yonago College, 4448 Hikona, Yonago, Tottori 683-8502, Japan*
[f]*Plasma Science and Fusion Center, Massachusetts Institute of Technology, 77 Massachusetts Avenue, NW17, Cambridge, MA 02139, U.S.A.*
[g]*Geodynamics Research Center, Ehime University, Matsuyama, Ehime 790-8577, Japan*

Keywords: Superconductivity, High pressure, Chalcogenide

**Abstract**

Layered ternary transition-metal chalcogenides have been focused as a vein of exploration for superconductors. In this study, TiGeTe$_6$ single crystals were synthesized and characterized by structural and valence state analyses and electrical transport measurements. The transport properties were measured under various pressures up to 71 GPa. The activation energy gets smaller as the applied pressure increases, and a signature of a pressure-induced metallization was observed under around 8.4 GPa. Under 13 GPa, pressure-induced superconductivity was discovered in this compound for the first time, with successive drops at 3 K and 6 K in the resistance, indicating the presence of multiple superconducting transitions. The superconducting transition temperature kept increasing as we further applied the pressure to the TiGeTe$_6$ single crystal in the performed pressure range, reaching as high as 8.1 K under 71 GPa.

# 1. Introduction

Layered transition-metal chalcogenides provide broad applications such as spintronics [1], nonreciprocal optics [2], gas sensors [3], thermoelectric materials [4], and monolayer physics [5]. Various parameters can tune their 2D structure to study the link between different electronic orders. Binary layered transition-metal chalcogenides have been focused as a vein of exploration for superconductors using various approaches such as chemical doping [6], electric field gating [7], and high pressure [8]. A rapid growth method for a high-quality single crystal of superconducting transition-metal chalcogenides has been developed via solvent evaporation [9]. Most recently, pressure-induced superconductivity at 6.5 K under 57 GPa and 6.6 K under 60 GPa were discovered in the ternary layered chalcogenides $ZrGeTe_4$ and $HfGeTe_4$, respectively [10]. Further exploration of ternary layered chalcogenides is expected. The synthesis of $(Zr,Hf)GeTe_4$ single crystal has been first shown in 1993 [11]. Although $TiGeTe_6$ was also synthesized in the literature, there is no further report until now.

The pressure-induced superconductivity on $(Zr,Hf)GeTe_4$ were discovered by a data-driven approach based on high-throughput computation using a database and density functional theory (DFT) calculations [10,12]. A flat band feature near the Fermi level ($E_F$) is used for the screening conditions. This is because the flat band induces a high density of state (DOS) near the $E_F$, preferable for the emergence of superconductivity. To achieve high superconducting transition temperature ($T_c$), the energy level and shape DOS have to be optimized, as has been intensively studied recently in high-$T_c$ hydrides [13] in the context of partial substitution of a host metal. In addition to such a partial substitution of constituent atoms, an application of high pressure is also a quite powerful tool to modify DOS, which can sometimes make insulators with narrow bandgap to metals and even induces superconductivity. According to first-principles calculations for $TiGeTe_6$, the DOS is expected to possess a sharp peak structure like the case of a van Hove singularity (vHs) near $E_F$ [11]. Moreover, a narrow bandgap of 0.2 eV is expected on $TiGeTe_6$ [11], a preferred feature for exploring the pressure-induced superconductors [12].

In this study, we report the first observation of pressure-induced superconductivity on $TiGeTe_6$ single crystal. The crystal structure, compositional ratio, and valence state were evaluated by single-crystal x-ray diffraction (XRD), energy-dispersive X-ray spectroscopy (EDX), and X-ray photoelectron spectroscopy (XPS), respectively. The electrical transport measurements were conducted by a diamond anvil cell (DAC) with boron-doped diamond electrodes [14-16].

# 2. Experimental

Single crystals of $TiGeTe_6$ were grown by referring to the literature [11] with slight modifications. Starting materials of Ge, Te, and Ti were put into an evacuated quartz tube in an atomic ratio of Ti : Ge : Te = 1 : 2 : 6. The excess Ge is considered to be acted as a flux to accelerate the crystal growth. The sample was heated to 650ºC, kept at 650ºC for 20 hours, heated to 900ºC, kept at 900ºC for 50 hours, slowly cooled to 500ºC for 50 hours, and naturally cooled to room temperature. The longer holding time and slower cooling rate give the larger crystals.

The obtained products were characterized by a chemical composition analysis using a scanning electron microscope (SEM) equipping EDX (JSM-6010LA, JEOL), single crystal XRD using XtaLAB mini (Rigaku) with Mo Kα radiation ($α$ = 0.71072 Å). The SHELXT on the WinGX software and ShelXle [17-19] were used to solve and refine the crystal structure of the product. For visualization of the crystal structure, VESTA software [20] was used. XPS has been performed using the AXIS-ULTRA DLD (Shimadzu/Kratos) with monochromatic Al Kα X-ray radiation ($hv$ = 1486.6 eV), under a pressure of the order of $10^{-9}$ Torr. In XPS, the background signals were subtracted by an active Shirley method on COMPRO software [21]. The binding energy of the sample was obtained by measuring a valence band spectrum of Au reference in the same chamber.

Resistance-temperature (*R-T*) measurements in the range of 300 to 1.8 K on TiGeTe$_6$ were measured with two settings. We used a conventional four-probe method for the measurement at ambient pressure using Au wires and silver paste painting. On the other hand, for the measurements under pressure up to 71 GPa, we used an originally designed DAC with the boron-doped diamond electrodes and the undoped diamond insulating layer [16]. As a configuration of DAC, we used a combination of a nano-polycrystalline (NPD) box-type anvil [22] and a single-crystal culet-type anvil with 300 μm culet, as shown in fig. S1. A single crystal sample was put on the boron-doped diamond electrodes of the NPD anvil, and the resistance was measured by a four-probe method. The suitable probe-current for the resistance evaluation was determined from the current-voltage relationship before the measurements. A sample chamber was made by a stainless-steel (316L) gasket with a drilling hole of a diameter of 200 μm. An undoped diamond insulating layer electrically separated the gasket and electrodes. A pressure-transmitting medium of cubic boron nitride was used with the ruby powders as a manometer. The pressure of the sample chamber was determined by a peak shift of ruby fluorescence [23] and Raman spectrum of diamond anvil [24] by an inVia Raman Microscope (RENISHAW). A temperature control, resistance measurements, and a magnetic field application were conducted using a physical properties measurement system (Quantum Design).

**3. Results and discussion**

Figure 1 (a) shows an optical microscope image of the obtained sample. The needle-like crystals with a typical length of ~1 mm were grown from the rock-shaped precursors with excess Ge. The crystal exhibited a well-cleavage feature along the longitudinal direction as presented in a typical SEM image of fig. 1 (b), indicating a Van der Waals-type layered structure. The compositional ratio is determined to Ti : Ge : Te = 1 : 0.98(2) : 6.10(3) as an average value of EDX on the three analysis points, consistent with a nominal composition of TiGeTe$_6$. Single-crystal XRD analysis was performed by putting a grown crystal on a glass capillary, as shown in an inset photo of fig. 1 (c). The refinement for TiGeTe$_6$ converged to $R_1$ = 11.32 %. The obtained TiGeTe$_6$ crystallizes in the *C* 2/*m* space group of monoclinic structure with lattice parameters $a$ = 13.970(8) Å, $b$ = 3.875(2) Å, and $c$ = 17.409(8) Å, and $α = γ = 90°$, $β = 104.94(3)°$. A significant decrease of the $R_1$ values by varying the site occupancy in Ti, Ge, and Te was not observed, suggesting no deficiencies in the analyzed TiGeTe$_6$ single crystal, unlike the case of other transition metal chalcogenide HfGeTe$_4$ [25]. Figure 1 (c) shows a schematic image of the crystal structures for TiGeTe$_6$ based on the single crystal structural analysis. The detail of the crystallographic data for TiGeTe$_6$ is shown in table S1.

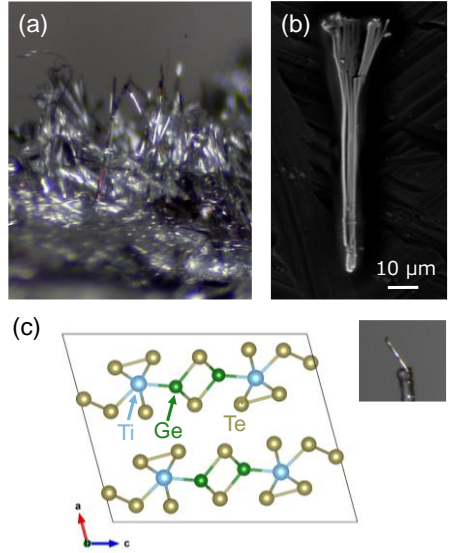

**Figure 1.** (a) Optical image of the obtained TiGeTe$_6$. (b) Typical SEM image of TiGeTe$_6$. (c) Schematic pictures of the crystal structures of TiGeTe$_6$ based on the single crystal structural analysis. The inserted optical image is a picked-up sample on the glass capillary for the single crystal XRD measurement.

An XPS analysis investigated the valence states of composed elements in the obtained TiGeTe$_6$ single crystals. Before acquiring the XPS spectra, the sample surface was cleaned by irradiation of the Ar gas cluster ion beam (GCIB) [26]. The beam energy, the mean size of one cluster, the scanning area, the beam current, and the irradiation time were 20 keV, 1000 atoms, 1 mm$^2$, 20 nA, and 5 min, respectively. Figure 2 (a) shows the valence band spectra near $E_F$ from TiGeTe$_6$ and Au as a reference for calibration of $E_F$ position. There is no band edge at the $E_F$ on the TiGeTe$_6$ spectrum, suggesting an insulating characteristic. Figure 2 (b) shows the Ti 2p spectrum splitting into two peaks by spin-orbit coupling, at 454.9 and 461.0 eV, corresponding to 2p$_{3/2}$ and 2p$_{1/2}$ of Ti$^{2+}$ [27]. Figure 2 (c) displays the Ge 3p core-level spectrum exhibiting the peaks located at 122.6 and 126.4 eV, which can be assigned to Ge 3p$_{3/2}$ and 3p$_{1/2}$, respectively, indicating Ge$^{4+}$ [28]. As shown in Te 3d spectra of fig 2 (d), Te 3d core-level spectrum presents Te 3d$_{5/2}$ and 3d$_{3/2}$ peaks at 572.7 and 583.2 eV, respectively, suggesting a valence state of Te$^{2-}$ [29]. All photoelectron spectra indicate no mixed-valence state in the composing elements.

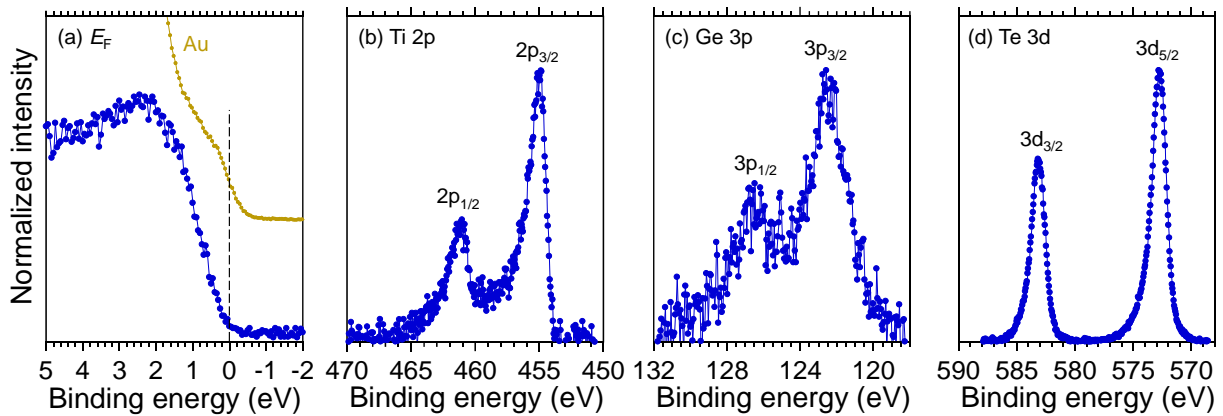

**Figure 2.** XPS spectra of (a) $E_F$, (b) Ti 2p, (c) Ge 3p, and (d) Te 3d acquired from TiGeTe$_6$ single crystals.

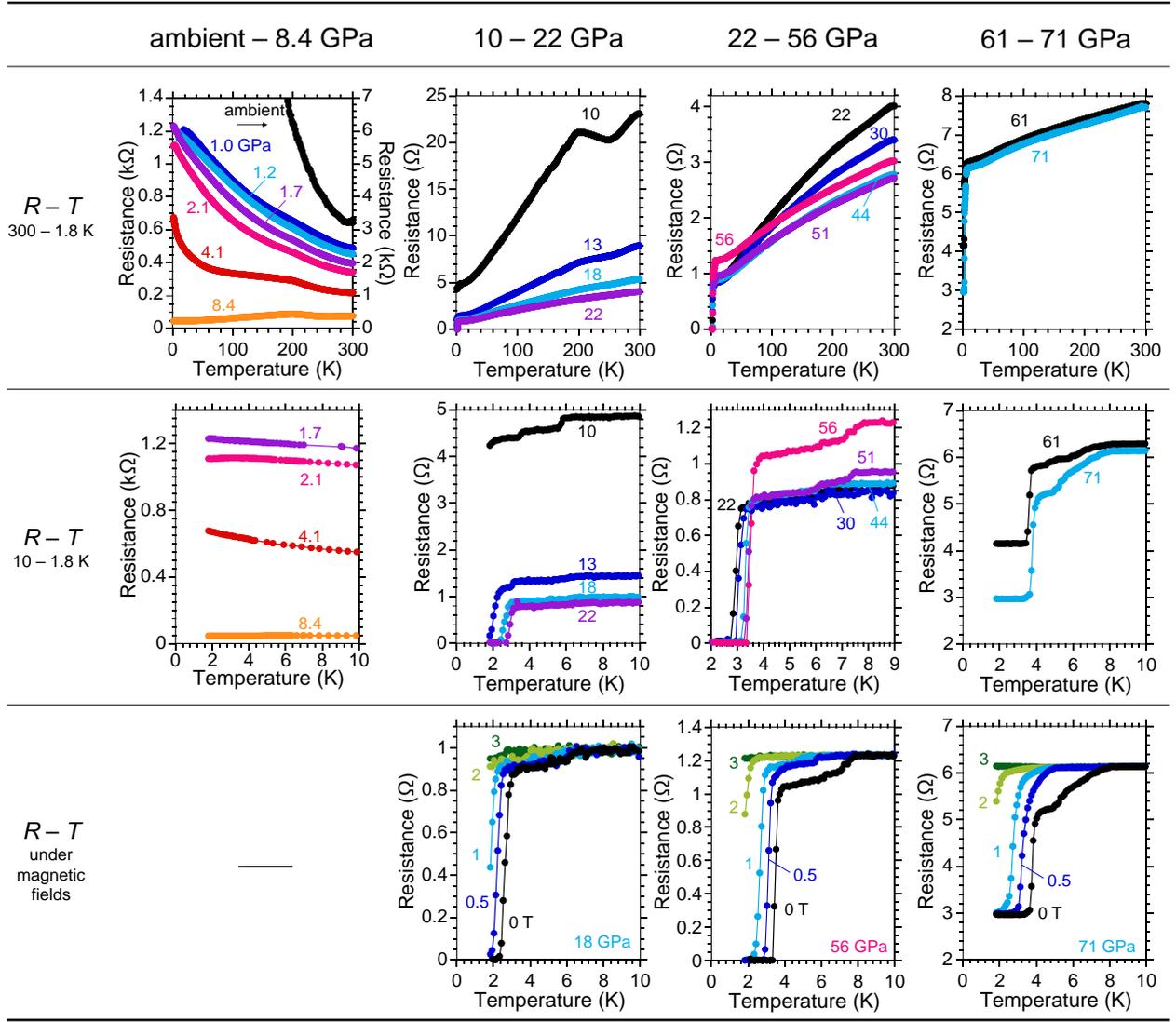

**Figure 3. Temperature dependence of resistance under various pressures in a TiGeTe$_6$ single crystal. The data is separated by a pressure region such as ambient-8.4 GPa, 10-22 GPa, 22-56 GPa, and 61-71 GPa, for reader's visibility.**

Figure 3 summarizes the temperature dependence of resistance under various pressures in a TiGeTe$_6$ single crystal. The data is separated by a pressure region such as ambient-8.4 GPa, 10-22 GPa, 22-56 GPa, and 61-71 GPa, to avoid excessive data overlap. At ambient pressure, TiGeTe$_6$ shows a negative slope of the *R-T* curve, suggesting an insulating nature. The Arrhenius relationship of $R=R_0 \times \exp(E_a/k_B T)$ was used to evaluate a pressure-induced metallization as shown in fig. 4 (a), where $R_0$ is the residual resistance, $E_a$ is the activation energy, $k_B$ is the Boltzmann constant, and $T$ is the temperature. The fitting was performed in the range of linear region of ln(*R*) versus 1/*T*. The estimated activation energy at ambient pressure is around 38 meV. By applying pressure up to 8.4 GPa, an absolute value of resistance is drastically decreased. The activation energy is linearly reduced against the applied pressure and reaches zero at 8.4 GPa, as shown in Fig. 4 (a), indicating the presence of insulator-to-metal transition upon pressure.

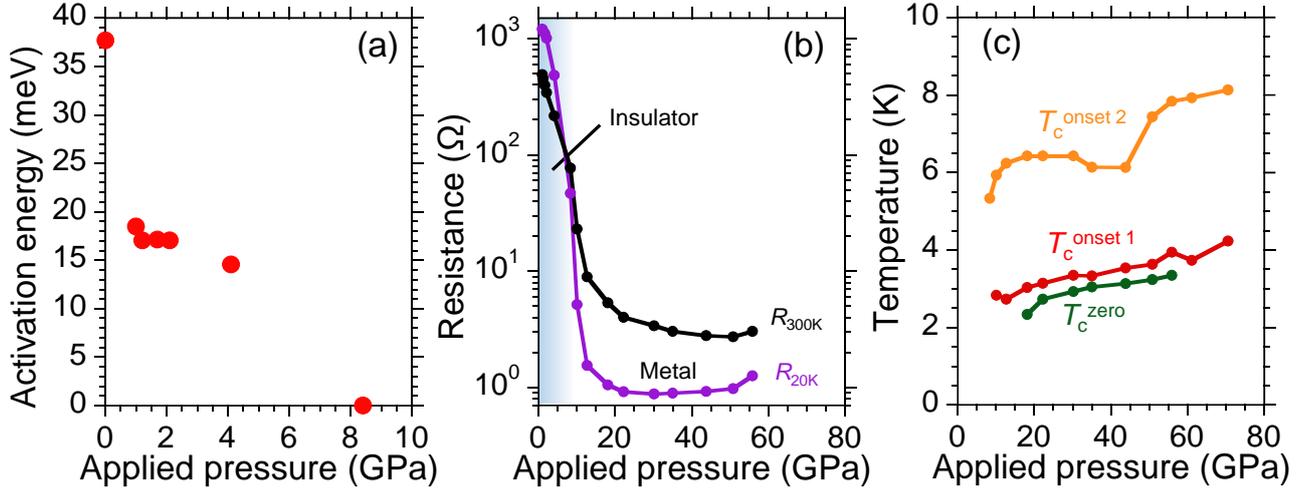

**Figure 4.** The relationship between various measurement values and the applied pressure on TiGeTe$_6$ single crystal. (a) Activation energy from Arrhenius plots, (b) electrical resistance at 300 K and 20 K, and (c) $T_c$s.

In the pressure range of 10-22 GPa, the TiGeTe$_6$ shows a sudden drop of resistance at 3 K with a tiny drop at 6 K. The $R$-$T$ curves were measured under various magnetic fields as a bottom row of fig. 3. Both drops of resistance at 3 K and 6 K are gradually suppressed by increasing the magnetic field, suggesting that both drops of resistance are originating from superconductivity with different $T_c$ values. Because a non-hydrostatic pressure medium of the cubic boron nitride is used, a hydrostaticity is low, as seen in the peak broadening of ruby fluorescence in Fig. S2 (a). Pressure distribution in the sample space is also large (see Fig. S2 (b)). The coexistence of two phases is possibly originated in the pressure distribution. The magnetic measurements under high pressure are expected to discuss the superconducting volume fraction as a further investigation. Moreover, a hump of the $R$-$T$ curve can be seen at around 200 K under 1.0 to 13 GPa. The hump temperature is almost constant against the increase of pressure. Such hump is found in various metal chalcogenides as a sign of charge density wave (CDW) [30]. By increasing the pressure above 18 GPa, a zero resistance appears instead of suppression of the hump. This fact implies that the suppression of the hump is preferred for superconductivity. A similar trend is reported on 1$T$-TiSe$_2$ in a previous study, where it has been revealed that the CDW state was suppressed by applying pressure, and then the superconductivity appeared [8]. The coexistence of superconductivity and CDW is also a hot topic, as seen in TaS$_2$ [31], NbSe$_2$ [32], HfTe$_3$ [33], and so on.

Under the pressure region of 22-56 GPa, the drop of resistance on the higher $T_c$ component became more prominent, indicating an enhancement of the superconducting current path by applying pressure. The $T_c$ value continues to increase up to 56 GPa. The zero resistance could not be observed above 61 GPa, which often indicates that the anvil is getting damaged and causing electrodes error. Despite, the onset $T_c$s has been successfully measured up to 71 GPa. The diamond anvil has been eventually broken while increasing the pressure above 71 GPa, unfortunately.

Figure 4 (b) shows a relationship between the electrical resistance and the applied pressure on TiGeTe$_6$. At the low-pressure region, the resistance at 20 K is higher than that at 300 K, indicating an insulating nature. By applying pressure, the resistance is drastically decreased more than three orders of magnitude. The resistance at 20 K becomes lower than that at 300 K at 8.4 GPa. Such behavior of resistance is a clear indication of the presence of a pressure-induced insulator-metal transition. Figure 4 (c) presents an applied pressure dependence on $T_c^{zero}$, $T_c^{onset1}$, and $T_c^{onset2}$ corresponding to a $T_c$ with zero resistance, lower $T_c$, and higher $T_c$, respectively. The $T_c^{onset2}$ reached as high as 8.1 K at 71 GPa. We have no clear answer for the origin of multiple $T_c$s, however, as a future investigation, it can be addressed by detailed structural analysis under high pressure using a synchrotron XRD and theoretical computation. Such research can contribute to an investigation for a relationship between superconductivity and CDW-like behavior.

## 4. Conclusion

Layered ternary transition-metal chalcogenide TiGeTe$_6$ that possesses a narrow band gap and high DOS near the $E_F$ was successfully synthesized in a single crystal. Pressure-induced metallization was observed at 8.4 GPa via the electrical transport measurements under high pressure. The emergence of superconducting transition was first discovered in TiGeTe$_6$ under 13 GPa. The maximum superconducting transition temperature reached 8.1 K at 71 GPa. This result would stimulate the further exploration of superconductivity in ternary transition-metal chalcogenides.

**Acknowledgment**

This work was partly supported by JST CREST (Grant No. JPMJCR16Q6), JST-Mirai Program (Grant No. JPMJMI17A2), JSPS KAKENHI (Grant No. 19H02177, 20H05644, and 20K22420). Fabrication of the diamond electrodes was partly supported by NIMS Nanofabrication Platform in Nanotechnology Platform Project sponsored by the Ministry of Education, Culture, Sports, Science and Technology (MEXT), Japan. The high-pressure experiments were supported by the Visiting Researcher's Program of GRC. The nano-polycrystalline diamond was synthesized by Toru Shinmei of GRC. RM would like to acknowledge the ICYS Research Fellowship, NIMS, Japan.

Table S1. Crystallographic data for TiGeTe$_6$.

| | |
|---|---|
| Structural formula | TiGeTe$_6$ |
| Formula weight | 886.09 |
| Crystal dimensions (mm) | 0.02×0.17×0.01 |
| Crystal shape | Needle |
| Crystal system | Monoclinic |
| Space group | *C* 2/*m* (No. 12) |
| *a* (Å) | 13.970(8) |
| *b* (Å) | 3.875(2) |
| *c* (Å) | 17.409(8) |
| *β* (°) | 104.94(3) |
| *V* (Å$^3$) | 910.6(8) |
| *Z* | 8 |
| $d_{calc}$ (g/cm$^3$) | 12.927 |
| Temperature (K) | 293 |
| *λ* MoK*α* (Å) | 0.71073 |
| *μ* (mm$^{-1}$) | 45.862 |
| Absorption correction | Empirical |
| $θ_{max}$ (°) | 27.462 |
| Index ranges | -17<*h*<17, -4<*k*<3, -22<*l*<16 |
| Total reflections | 1703 |
| Unique reflections | 1026 |
| Observed [*I* ≥ 2*σ*(*I*)] | 561 |
| No. of variables | 25 |
| *R*1 [*I* ≥ 2*σ*(*I*)] | 0.1267 |
| *R* (all reflections) | 0.1692 |
| *wR*2 [*I* ≥ 2*σ*(*I*)] | 0.2960 |
| *GOF* on $F_o^2$ | 1.319 |

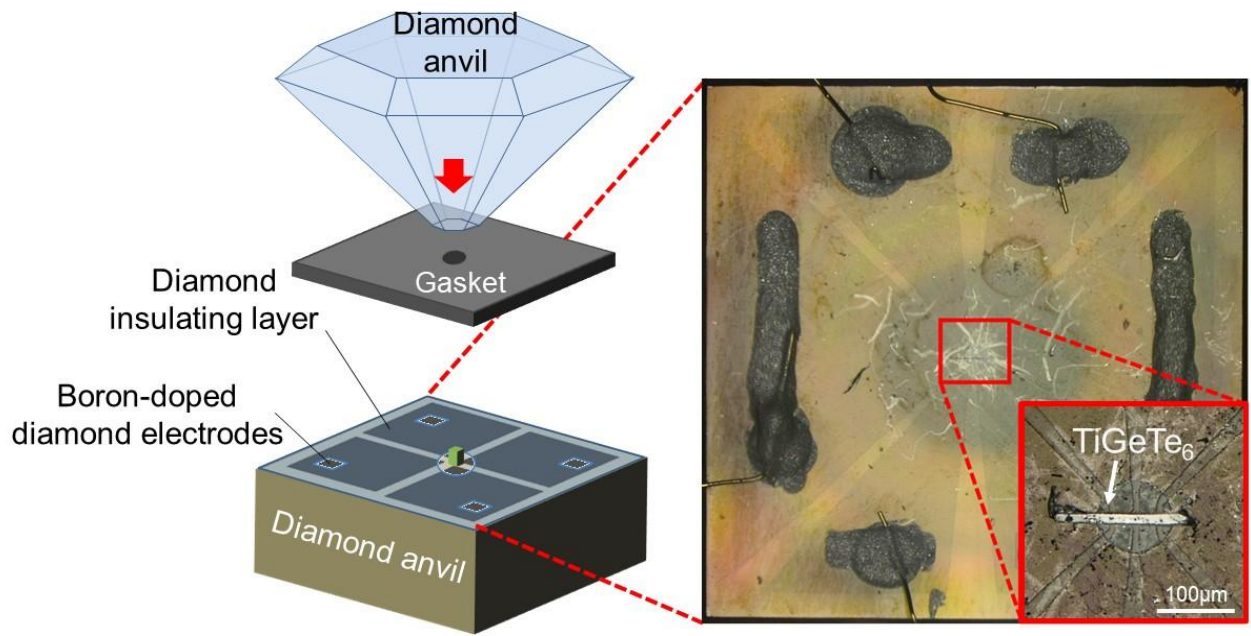

**Figure S1. Layout of diamond anvil cell for high-pressure experiments**

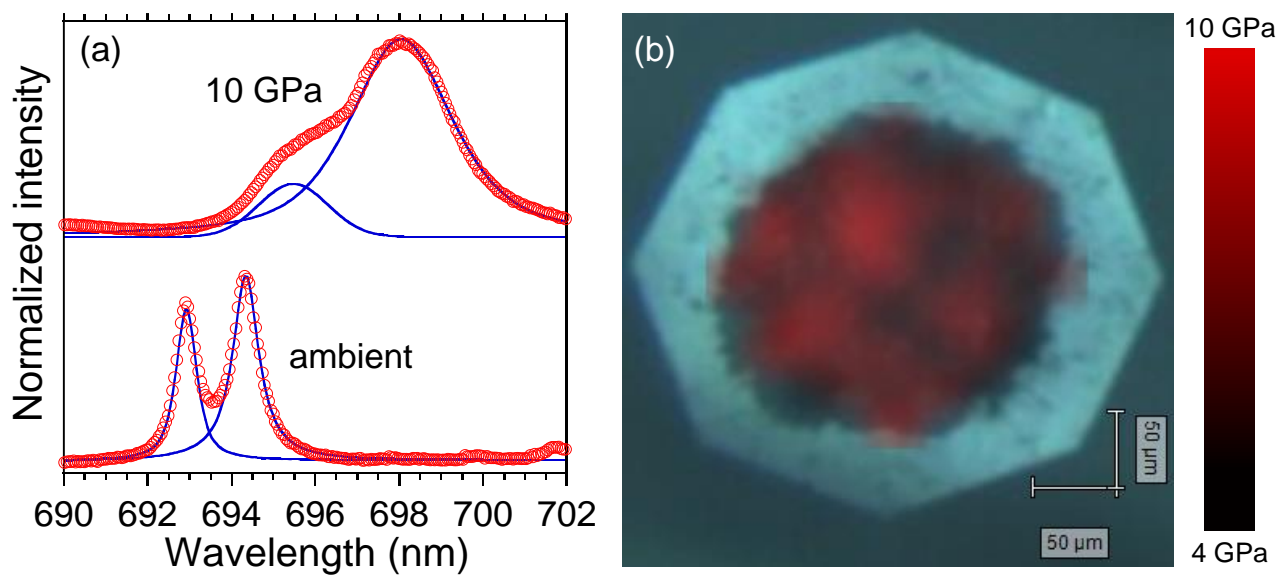

**Figure S2. (a) Comparison of ruby fluorescence under ambient and high pressure of 10 GPa. (b) Pressure distribution in the sample space of DAC under 10 GPa.**